\documentclass{zermatt}
\usepackage{graphicx}
\usepackage{natbib}
\usepackage[T1]{fontenc}
\usepackage[utf8]{inputenc}
\usepackage[english]{babel}
\usepackage{pdflscape}
\usepackage{url}

\newcommand{\CII}{[C\,{\sc ii}]}

\newcommand{\OI}{[O\,{\sc i}]}

\begin{document}
\title{Current and Future Space and Airborne Observatories for ISM Studies} 
\runningtitle{Schulz \& Meixner: Current and Future Space/Airborne Observatories for ISM Studies}

\author{B.~Schulz}
\address{Deutsches SOFIA Institut, University of Stuttgart, Pfaffenwaldring 29, 70569, Stuttgart, Germany,\\ \email{bschulz@dsi.uni-stuttgart.de}}
\sameaddress{, 2}

\author{M.~Meixner}
\address{SOFIA Science Center, NASA Ames Research Center, Moffett Field, CA 94045, USA,\\ \email{mmeixner@usra.edu}}

\begin{abstract}
A tremendous amount of radiation is emitted by the Interstellar Medium in the mid- and far-infrared (3-500~$\mu$m) that represents the majority of the light emitted by a galaxy.
In this article we motivate ISM studies in the infrared and the construction of large specialized observatories like the Stratospheric Observatory For Infrared Astronomy (SOFIA), which just concluded its mission on a scientific high note, and the newly launched James Webb Space Telescope (JWST) that just begun its exciting scientific mission.
We introduce their capabilities, present a few examples of their scientific discoveries and discuss how they complemented each other. We then consider the impact of the conclusion of SOFIA for the field in a historic context and look at new opportunities specifically for far-infrared observatories in space and in the stratosphere.

\end{abstract}
\maketitle

\section{The Interstellar Medium}

The Interstellar Medium (ISM) constitutes the reservoir of matter, that was and still is turned into stars and planets and gave also rise to the existence of our own solar system and our world. 
If its study wasn't already interesting for just that reason, there are many complex processes that impact it chemically as well as energetically.

The ISM is being enriched with heavier elements by the more massive stars in their late evolutionary phases, but also diluted by the influx of extragalactic matter. 
Feedback from the different phases of stellar life, but also cosmic rays and AGN inject energy, which is released by emission in many atomic and molecular lines  (\CII, \OI, C$_2$H$_2$, H$_2$O, PAHs, etc.) as well as thermal emission by different kinds of dust.
Even though the general scenario of star formation is reasonably well understood, the details of the complex interplay of stellar radiation, gravitation, turbulence and magnetic fields, that determine the timescales and the interstellar mass function, are not.

\begin{figure}[htp]
      \begin{minipage}{0.50\textwidth}
	\includegraphics[width=\textwidth]{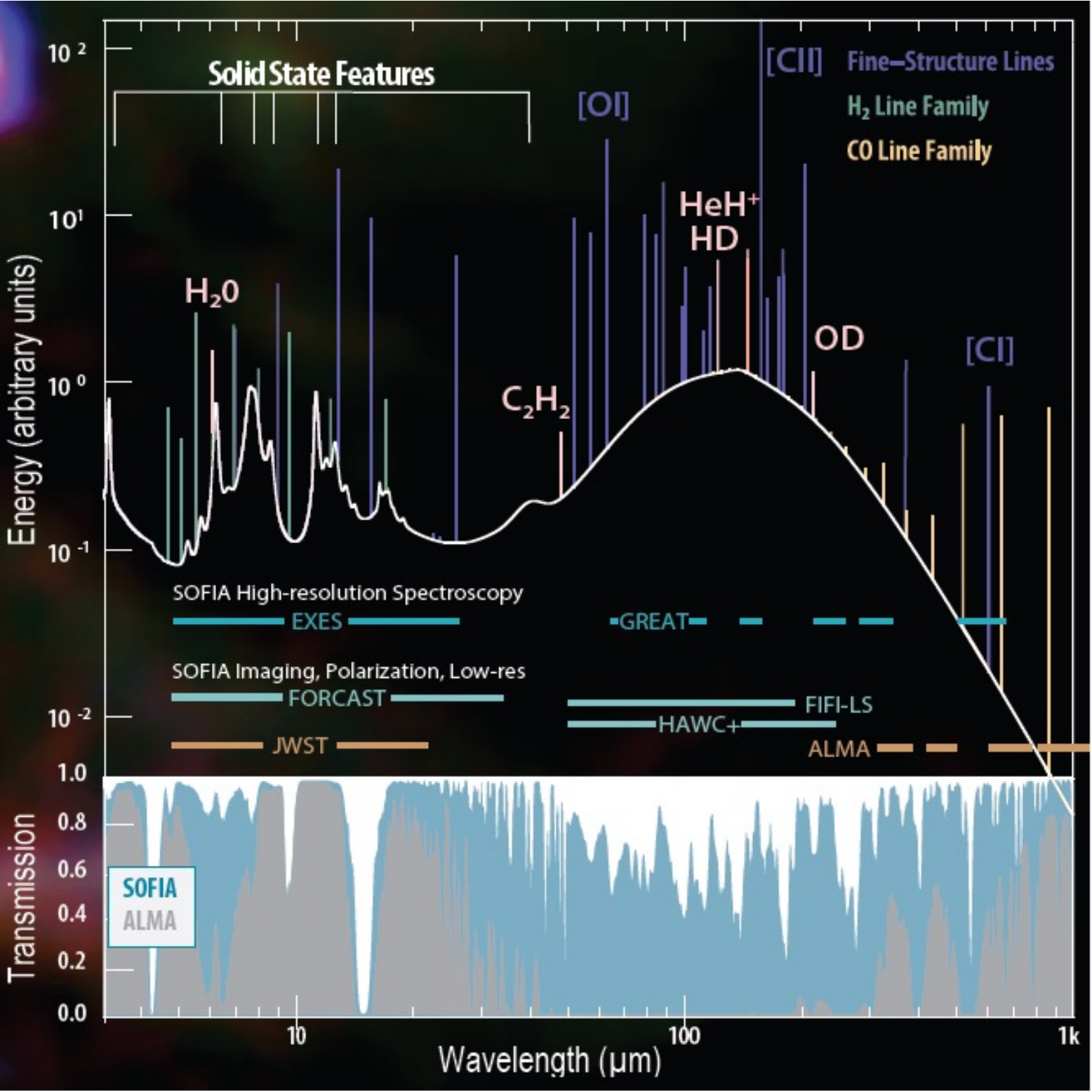}
      \end{minipage}
      \hspace{0.5cm}
      \begin{minipage}[t]{0.4\textwidth}
	\caption{The Spectral Energy Distribution (SED) of the interstellar medium from mid- to far-infrared wavelengths (top) and the corresponding transmission spectra of the Earth atmosphere at the operating altitudes of ALMA and SOFIA (bottom).}
	\label{Meixner:fig:ismsed}
      \end{minipage}
\end{figure}

A large number of these lines as well as the peak of the thermal emission are located in the mid- to far-infrared (MIR 3-30$\mu$m, FIR 30-300$\mu$m) wavelength range as illustrated in  Fig.~\ref{Meixner:fig:ismsed} (top), making this portion of the electromagnetic spectrum key to studying the ISM and a multitude of related scientifically interesting phenomena.
However, this is also a rather difficult spectral range to observe as shown in the lower part of Fig.~\ref{Meixner:fig:ismsed}, which illustrates the atmospheric transmission at the levels of the Atacama Large Millimeter/submillimeter Array (ALMA) and the Stratospheric Observatory for Infrared Astronomy (SOFIA). Telluric water vapor and ozone leave only certain windows in the MIR and sub-millimeter ranges, while the FIR is effectively unobservable from the ground and requires observatories in the stratosphere or in space.

\section{JWST}
Launched at the end of 2021, the James Webb Space Telescope (JWST) provides access to the full MIR spectrum in space since its first science data were released in July 2022. 
Its high spatial resolution, similar to that of Hubble in the visible spectrum, and its access to PAH emission as well as the ro-vibrational lines of molecular hydrogen and water, make JWST an excellent probe of star formation regions. 
JWST images of 30 Doradus, aka the Tarantula Nebula, show the stars, molecular hydrogen and PAHs with NIRCam (0.6-5~$\mu$m)  and warm dust and PAHs with MIRI (4.9 to 28.8~$\mu$m). 
These kind of maps provide an unprecedented amount of detail at those wavelengths and will play an important role in further investigating the hot and warm ISM. 
The spectroscopic capabilities of JWST are considerable as well, yet limited in terms of spectral resolution with R~$\approx 2700$ \citep{meixner:cite:Boeker2022} for NIRSpec and R~$\approx 1300$ to 3700 for MIRI \citep{meixner:cite:Wells2015}.
This is where SOFIA provided complementary high spectral resolution spectroscopy (R~$\approx10^5$) with EXES \citep{meixner:cite:Richter2018}, even though at lower spatial resolution and sensitivity.

\section{SOFIA}
\subsection{Importance and Successes}
With its five exchangeable scientific instruments (SIs), SOFIA filled nicely the large spectral gap in the FIR between JWST and ALMA and provided further complementary capabilities like high spectral resolution at JWST wavelengths with EXES and the ability to observe very bright sources with FORCAST, that filled in the overexposed areas in MIR maps of the Galactic Center region, made by Spitzer \citep{meixner:cite:Hankins2020}.
The recent discovery of water on the sunlit surface of the Moon by \cite{meixner:cite:Honniball2021} falls into that category as well.
The heterodyne instrument GREAT covers such important atomic fine structure lines as \CII~and \OI~at the highest spectral resolutions of up to R~$\approx 10^6$ and fills in the spectral gaps that are inaccessible for ALMA due to atmospheric extinction.
This enabled not only the discovery of new molecules in the ISM like Helium Hydride \citep{meixner:cite:Guesten2019}, but also very detailed kinematic studies e.g. feedback processes in Orion by \cite{meixner:cite:Pabst2019} that triggered a very successful SOFIA legacy program by \cite{meixner:cite:Schneider2020}.
When sensitivity became an issue and could be gained by sacrificing spectral resolution, in particular for extragalactic work, the FIFI-LS spectrometer provided a good alternative for observations of fine structure lines as shown by \cite{meixner:cite:Fadda2021}, \cite{meixner:cite:Spinoglio2022} or \cite{meixner:cite:Pineda2018}.
Last but not least, where very high sensitivity was required to reveal the peak of the cold dust emission of high redshift objects, HAWC+ provided the FIR imaging capability.
HAWC+, however, also provided an entire new dimension for ISM studies, that had before only briefly been available with ISO in the FIR.
Polarization mapping revealed the vectors of magnetic fields in the ISM thanks to the FIR emission of aligned elongated dust particles. Many publications sparked a lot of new observational as also theoretical interest in this previously rather dormant field \citep{meixner:cite:Pillai2020, meixner:cite:Lopez-Rodriguez2021, meixner:cite:Zielinski2021}.

\subsection{Mission Success and Conclusion}
In the face of the tremendous scientific successes of this true ISM-Machine, the decision by NASA and DLR to end the SOFIA mission after only 9 observing cycles, is certainly very hard to understand.
Following the recommendations from the Flagship Mission Review from 2019, the project has transformed since then with a tremendous growth in science productivity as demonstrated in the SOFIA Status and Future Prospects Report \citep{meixner:cite:Rangwala2022}\footnote{This report was already prepared for NASA's Senior Review Process.}.
Annual publication rates for SOFIA have doubled over the past three years on topics ranging from the Earth to high-z galaxies \citep{meixner:cite:Schmelz2021}.

The Decadal Survey Astro~2020 recommended to NASA to terminate the SOFIA mission, which unfortunately was based on outdated ($>2$~years) and incorrect information\footnote{SOFIA science addresses 50\% of Astro~2020 key science questions, not 10\%.}.
NASA holds Astro 2020 recommendations as superior to Senior Review process results and hence removed SOFIA from the Senior Review.\footnote{This avoided potentially ending up with two contradicting recommendations.}
Arguments that SOFIA's science productivity was insufficient can be easily refuted by comparing the observing time that is spent on average per refereed publication to that of Herschel. Eight years after launch Herschel had provided about 23,500 hours of observing time and produced 2,145 publications, resulting in $\approx$11~h$/$paper. SOFIA with 3458 hours and 330 publications after 8 years since achieving full science operational capability in 2014 results in very similar 10.5~h$/$paper.

Fortunately the last year was particularly productive in terms of observations, so there is a considerable amount of science data in the IRSA archive. As there is only a minimal post-operational phase of one year planned by NASA at this point, we hope DLR will provide the means to conduct data reprocessing also for the time before Cycle~5, advanced water vapor and pointing analysis and more comprehensive corrections, which are currently not included in the plans. In the next section we'll lay out that the time to the next FIR mission might be rather long. Already collected FIR photons might thus be even more valuable for astronomy and funds for maximizing their scientific usability will be well spent.

\section{Future Far-Infrared Observatories}

\subsection{History and Guidance}
\begin{landscape}
\begin{figure}[htp]
    \centering
	\includegraphics[scale=.9]{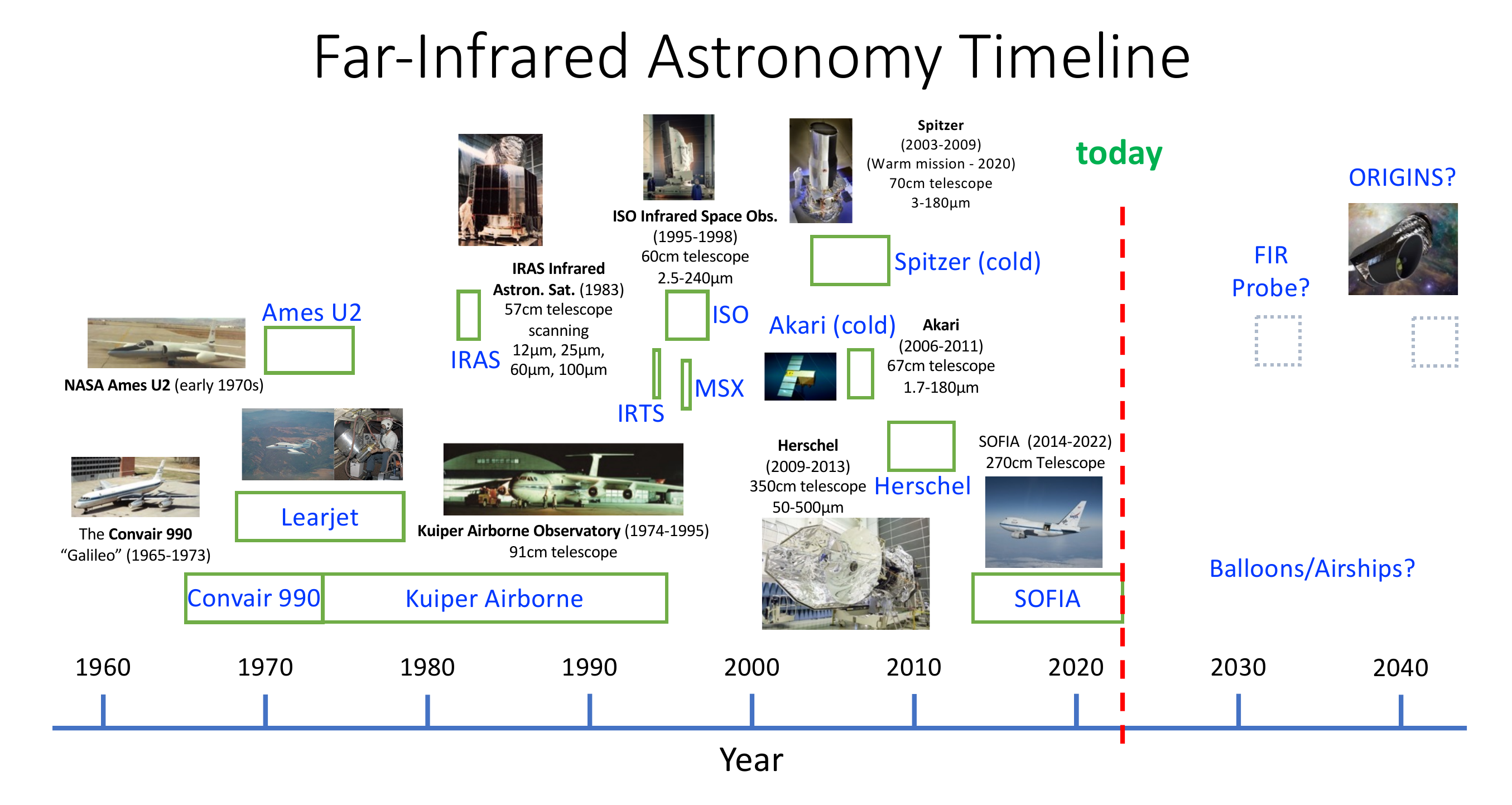}
	\caption{The historic timeline of FIR stratospheric- and space-missions: The green boxes represent the intervals when the respective observatories, annotated in blue, were conducting science operations. The dashed vertical line, marked as "today", represents the time of this conference in 2022, which happened to coincide with the final scientific flight of SOFIA. To the right extends the future until the 2040s, illustrating the lack of plans for FIR observatories, that historically, at least for space missions, took about one to two decades from inception to launch.}
	\label{Meixner:fig:firhistory}
\end{figure}
\end{landscape}

Fig.~\ref{Meixner:fig:firhistory} illustrates the history of FIR astronomy by showing the operational phases of all major observatories as green boxes, starting in the sixties until today and the current outlook towards 2045. Up to today, there was an almost continuous capability to supply astronomers with current FIR observations except for the few years between ISO and Spitzer. With the sudden cancellation of SOFIA, which was originally scheduled to continue until 2034, and the cancellation of SPICA by ESA in 2021, the opportunities for FIR data collection have become sparse.
In \cite{meixner:cite:Rangwala2022} Page~4, a traceability matrix can be found, that links Astro 2020 science questions to key measurements in the MIR and FIR, that could have been performed with SOFIA.
This list should still be useful as a collection of science requirements for the design of future stratospheric- and space-observatories.

\subsection{New Opportunities in Space}
Even though Astro~2020 recommended the cancellation of SOFIA, it acknowledged the importance of the FIR spectral region for astrophysics and recommended the launch of a Probe space mission for 2030 that will specialize either in FIR- or X-ray- astronomy.
NASA followed this up by issuing an announcement of opportunity and a proposal deadline of October 2023, a downselection end of 2025, a cost cap of 1B$\$$ excluding the launcher and a launch date not later than 2032 \citep{meixner:cite:NASA-AO2022}.
If history is a guide, such a schedule is highly optimistic. In reality a launch might rather be expected in the mid 2030s, not to mention that continuing the SOFIA mission until its planned end would have cost substantially less, especially when taking into account the launcher as well.

Given that the X-ray community is also competing for another opportunity, it is everything but a done deal that NASA's probe mission will be dedicated to the FIR.
If that doesn't happen, then also the dream of a more ambitious true observatory for the FIR such as ORIGINS in the 2040s \citep{meixner:cite:Meixner2019} may become unrealistic with observational FIR astronomy having lost a lot of its expertise by then.

Therefore at this point it is quite important for the FIR community to look towards the future which at least in space will be the Probe mission. There are four mission proposals for the FIR named PRIMA (PI, Jason Glenn)\footnote{\url{https://prima.ipac.caltech.edu}}, SPICE (PI Lee Mundy)\footnote{\url{https://asd.gsfc.nasa.gov/spice/index.html}}, FIRSST (PI Asantha Cooray) and SALTUS (PI Chris Walker), which were presented at the IR Astrophysics Workshop 2022 in Colorado.  The concepts comprise more traditional space observatories with cold telescopes like PRIMA and FIRSST, and more unusual ones like the interferometer SPICE or the large inflatable telescope concept SALTUS. Details as presented at the workshop are available at the workshop website \citep{meixner:cite:Irstig2022}.

\subsection{Stratospheric Opportunities}
In the meantime the FIR community should also investigate other opportunities to reclaim a permanent capability in that part of the spectrum. 
This will in particular enable more time dependent FIR astronomy, that we consider being still in its infancy.
The fairly short life spans of FIR missions so far have been a hindrance while time-domain astronomy has really taken off in other parts of the electromagnetic spectrum.
The astrophysical community should investigate the available potential in the FIR.

SOFIA was likely the last airplane observatory, and future stratospheric platforms will probably be of the lighter-than-air category. 
Current balloon experiments are, however, rather short lived, extremely weather dependent with very few launch opportunities, can't stay in a particular region for long and have only a 50~\% survival rate upon landing. 
Such missions are still seen rather as serving technology maturation and the training of instrumentalists than being able to support serious general observatory type projects for the astronomical community.
This school of thought needs to change as better technologies become available that could address many of the shortcomings mentioned above.
Longer lived robotic stratospheric platforms with propulsion may also be interesting to a wider community including UV- and FIR-astronomy but also climate research and general Earth observation \citep{meixner:cite:Miller2014}.

\section{Conclusion}
Even though the end of SOFIA is a blow to FIR astronomy, the mission and its team have performed excellently and are concluding at peak performance with much data in the archive that await analysis and publication.
JWST is the observatory now to study the ISM in warm/hot conditions, while there will be new opportunities for observatories that can study the cold ISM from space or from the stratosphere.

%

\end{document}